\documentclass[conference]{IEEEtran}
\usepackage{amsmath}
\usepackage{amsfonts}
\usepackage{setspace}
\usepackage{array}
\usepackage{amssymb}
\usepackage{graphicx}
\usepackage{url}
\usepackage{multicol}
\usepackage{stfloats}
\usepackage[usenames]{color}
\usepackage{flushend}
\usepackage{booktabs}
\usepackage{changes}
\usepackage{eurosym}
\usepackage{rotating}
\usepackage{color}
\usepackage{subcaption}
\usepackage{caption}
\usepackage{epstopdf}

\begin{document}

\title{Popularity Evolution of Professional Users on Facebook}

\author{
\IEEEauthorblockN{
Samin Mohammadi\IEEEauthorrefmark{2},
Reza Farahbakhsh\IEEEauthorrefmark{2},
No\"{e}l Crespi\IEEEauthorrefmark{2}}
\IEEEauthorblockA{\IEEEauthorrefmark{2}Institut Mines-T\'el\'ecom, T\'el\'ecom SudParis, CNRS UMR 5157 SAMOVAR, France,\\
\{samin.mohammadi, reza.farahbakhsh, noel.crespi\}@it-sudparis.eu}
}

\maketitle

\label{sec:Abstract}
\begin{abstract}
Popularity in social media is an important objective for professional users (e.g. companies, celebrities, and public figures, etc). A simple yet prominent metric utilized to measure the popularity of a user is the number of fans or followers she succeed to attract to her page. Popularity is influenced by several factors which identifying them is an interesting research topic. This paper aims to understand this phenomenon in social media by exploring the popularity evolution for professional users in Facebook.
To this end, we implemented a crawler and monitor the popularity evolution trend of 8k most popular professional users on Facebook over a period of 14 months. The collected dataset includes around 20 million popularity values and 43 million posts. We characterized different popularity evolution patterns by clustering the users temporal number of fans and study them from various perspectives including their categories and level of activities. Our observations show that being active and famous correlate positively with the popularity trend.

\end{abstract} 


\section{Introduction}
\label{sec:Introduction}

In the fast-paced digital world, Online Social Networks (OSNs) have experienced a massive growth in their variety and usage over the past decade. These systems offer a huge opportunity for professional users (i.e. companies, politicians, celebrities, etc.) who aim to both attract new followers and interact better with them \cite{Farahbakhsh:prof_users}.
Facebook as the most popular OSN with more than one billion subscribers
defines a specific type of account for professional users, called \textsl{FanPages}\footnote{http://www.facebook.com/about/pages/}.

This type of account has several features that distinguish it from regular accounts. %different than the normal one.
If a user likes a page, it will be added to the interest list of the user's profile.
Professional users from various categories can create \textsl{FanPages} on Facebook as a means of interacting with their fans and customers.
Apart from the general static attributes such as the page description and category selected by the page owner, the main dynamic attribute for each page is
the number of fans ($N_f$) who have liked the page. % and are following it.
This metric is publicly available for each \textsl{FanPage} and considered as the main metric that shows the popularity of a \textsl{FanPage} \cite{nelson2012s}. Even in major political events such as US presidential election, the popularity metric in different social media is the main metric to compare different candidate success in their campaign. 
Several studies have emphasized the role of $N_f$ as a comparative and competitive metric for professional users.
Many of professional users are willing to spend a considerable amount of money to increase this value, even through unusual ways such as buying likes from \textsl{like farms} \cite{de2014paying}\cite{stringhini2012poultry}. The number of likes of a page has been found to be one of the most positive correlated features linking candidates' fan pages to the number of their votes in elections \cite{barclay2014political}\cite{giglietto2012if}.
Attracting Facebook fans is also used as a marketing strategy \cite{hollis2011value} and provides a metric to measure the return on social media investment \cite{hoffman2010can}.
We will use the term \textit{popularity} to refer to the number of likes of a page.
To the best of authors' knowledge, even though a number of papers have studied the popularity trends of content and posts \cite{de2012popularity} \cite{yu2011toward}, there is no study on evaluating the popularity evolution of users, especially by the focus on professional users.

This paper studies the temporal popularity evolution of professional users through their \textsl{FanPages} on Facebook and attempts to identify the factors that influence the popularity trends. 
The objectives pursued here are designed to answer the following research questions:

$(i)$ How does the temporal popularity of users vary overall and in accordance with users' business sector (Facebook pre-defined categories)? 
$(ii)$ What temporal patterns can be identified from the time-series $N_f$ of pages?

$(iii)$ What are the factors influencing the popularity trends? 

To answer the stated questions, an extensive list of the most popular professional users in terms of $N_f$ was selected and the required data collected by implementing advanced data collection tools.
Our dataset includes 8K of \textsl{FanPages} that have the highest number of fans validated by a third-party portal \textsl{Social Bakers}\footnote{http://www.socialbakers.com/}.

The main contributions of this study are:

i) The proposed methodology of monitoring the popularity evolution of professional users on Facebook in very micro level is novel which is applicable to different types of OSNs.

ii) Following the methodology, we classified the users in two main groups: First, fan-attractors who grew their $N_f$ by different patterns, and second, fan-losers, users with a noticeable drop in their popularity trend.

iii) We found several influential factors on the popularity trend of users. The activity level of users or being celebrity are positively correlated to the trend of the number of fans. 
The rest of this paper is organized as follows: We present related work in Section \ref{sec:related_work} followed by Section \ref{sec:dataset} describing the methodology and the dataset. Section \ref{sec:Overall_Analysis} represents a general overview of the popularity and its evolution. The model and results are discussed in Section \ref{sec:Clustering} 
 and finally Section \ref{sec:Conclusion} concludes this study.

\section{Related Work}
\label{sec:related_work}
One of the most well-studied aspects of social media is popularity \cite{szabo2010predicting} \cite{de2012popularity} because popularity has become one of the main utilities that is used in advertisements, marketing, and predictions \cite{de2014paying}.
The term 'popularity' refers to different metrics such as the number of likes, views, or votes that a page or a content receives \cite{szabo2010predicting} \cite{cvijikj2013online}.
Barclay \textit{et al.} \cite{barclay2014political} investigated the correlation between political opinions on Facebook and Twitter in the US presidential elections of 2012. They showed that the number of fans and the sentiment of comments are the most-correlated features to the candidates final votes. In another similar work, Barclay \textit{et al.} \cite{barclay2015india} demonstrated the number of likes of the Facebook \textsl{FanPage}s of the parties as a predictor of election outcomes with 86.6\% accuracy.

Meanwhile, a number of studies have focused on identifying the influential factors on attracting new fans and increasing users' engagement level \cite{leung2013attracting} \cite{jayasingh2015customer}.
Authors in \cite{lombardi2012social} performed an empirical study on a sample of posts created by different brands on their Facebook \textsl{FanPages}. They investigated, the impact of some factors such as emotion and testimonial presence. 
Cvijikj \textit{et al.} \cite{cvijikj2013online} analyzed the effects of content characteristics on user engagement in Facebook \textsl{FanPages}. They found that providing informative and entertaining content significantly increases the user's engagement level.
To enhance the number of likes and comments of a post, Vries \textit{et al.} \cite{de2012popularity} found that highly vivid and interactive posts like videos and questions can attract more likes and comments than other kinds of post. Pronschinske \textit{et al.} \cite{pronschinske2012attracting} studied the relationship between the attributes of Facebook pages and the number of page likes. They showed that being authentic by indicating a page as an official page and linking a website to a Facebook page as well as having more engagement in the posts of a page will attract more fans.

Simultaneously, many studies have tried to model and forecast popularity, specially for content \cite{szabo2010predicting}. Bandari \textit{et al.} utilized article features like source, category, and subjectivity to predict the popularity of an article on Twitter with 84\% accuracy. Lerman \textit{et al.} used a stochastic model to predict how popular a newly posted story will be based on the early reactions of Digg users \cite{lerman2010using}. In \cite{figueiredo2014improving} and \cite{Hu2014} researchers used temporal content features to predict the popularity of content by exploiting time series clustering techniques and linear regression methods. Different categories of features have been examined to predict the popularity of content \cite{cheng2014can} and in \cite{shulman2016predictability} temporal features are illustrated as the best predictors.

It is worth mentioning that several companies monitor Facebook \textsl{FanPages} activities and provide reports, by charging their customers, with general analysis for their clients. One of them that provides aggregated popularity results for single users, is \textsl{SocialBakers}. They claim that 
their services allow brands to measure, compare, and contrast the success of their social media campaigns with competitive intelligence. 
In summary, although few studies have looked to the different aspects of Facebook \textsl{FanPages}, but their focus were mostly for a small group of users. To the best of the authors knowledge none of the previous studies has specifically investigated the evolution of popularity in a large scale and for a long period. This paper is the first study that looks to this aspect for a list of 8K popular \textsl{FanPages} and also investigates the influential factors to the popularity evolution trends.

\section{Data Collection and Dataset}
\label{sec:dataset}

% Dataset Table
\begin{table}[t]
  \centering
  \caption{Dataset Characteristic}
  \vspace{-0.2cm}
\scalebox{0.8}
{
    \begin{tabular}{|l|c|} 
    \hline
    \multicolumn{1}{|c|}{\textbf{Attribute}} & \multicolumn{1}{c|}{\textbf{Value}} \\
    \hline
    Duration & 14 months\\
    Crawling Period & Sep'13 - Oct'14\\
  %  Start date of data collection  & September 2013 \\
  % End date of data collection  & October 2014 \\
    \#Sample per day & 6 snapshots (Q4h)\\
		\#Users (\#\textsl{FanPages}) & 7,875 \\
    \hline
    Total \#Samples in dataset & 20M samples\\
    Avg(\#Sample) per user & 1,298 samples \\
    Median(\#Sample) per user & 1,297 samples\\
%	  \hline
%    Avg(\#Valid days with at least 1 sample) per user & 327 days\\
%    Median(\#Valid days with at least 1 sample) per user & 336 days\\
    \hline
    Total \#Post in dataset & 43M posts \\
%	Avg(sum(\#User\_Posts over 14 month)) & 5,666 posts \\
%    Median(\#Post per user over 14 Months) & 1,559 posts \\
   Avg(\#User\_Post) per month  & 107 posts \\
    Median(\#User\_Post) per month & 24 posts\\
    \hline
    \end{tabular}%
		}
  \label{tab:dataset_table}%
	\vspace{-0.1cm}
\end{table}%

The objective of this study is to explore how the popularity of top professional Facebook \textsl{FanPages} evolves. To this end, we first selected 8K of the top Facebook \textsl{FanPages} based on their $N_f$ from the previously mentioned third-party application \textsl{Social Bakers} 
which ranks users based on the number of fans. 

In order to monitor the popularity evolution of the selected users and generate a time-series of their $N_f$ and of their activities; we implemented three crawlers as follows:
Firstly, we implemented a data collection tool that queries FB public API to collect the number of fans. The data collection is performed for the selected 8K users over a period of 14 months from September 2013 to October 2014. To have enough detail, the value of $N_f$ is recorded, every 4 hours (6 times per day). 
The second crawler collects the general information of users from their profile which includes detailed information such as their pre-defined categories , description of the page, etc.
The third crawler collects the activity (published posts) of users and its associated attributes on the period of our study.
A summary of dataset's main characteristics is presented in Table \ref{tab:dataset_table}.

\section{Evolution of Popularity}
\label{sec:Overall_Analysis}
Before clustering, we go through the analyzing aggregated popularity evolution of users to provide an insightful vision of the dataset. During the initial analysis, a group of users is identified who have a sudden and large peak in their $N_f$ in a very short period of time . By looking carefully to their data, we found that this peak reflects the impact of a newly announced service by Facebook, named \textsl{GlobalPage} \cite{facebook_globapage}. Facebook \textsl{GlobalPage} is a new page structure for big brands which are active across globe and have several separate pages with the same name but active in different languages and different locations.% (e.g. countries or continents).
These pages which formed almost 10\% of the dataset,  were excluded from it because their trend are not aligned with the aim of this study which is to identify real popularity trends and their effective factors.

\subsection{Popularity Analysis - In Overall}

\begin{figure}[t]
%\hspace{-0.2cm}
\begin{minipage}[b]{0.51\linewidth}
	\centering
%	\vspace{-0.5cm}	
	\centering
	\includegraphics[width=\textwidth]{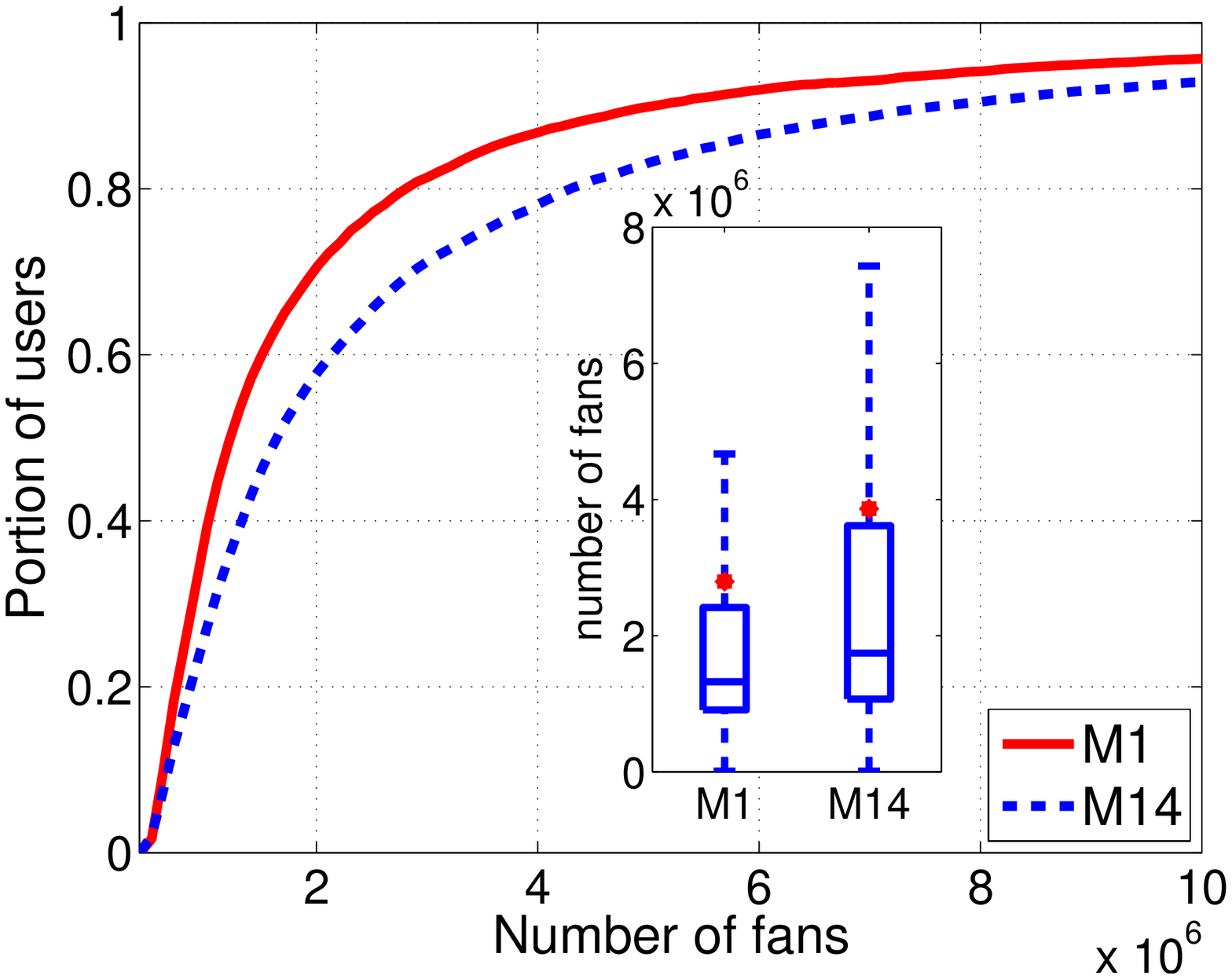}	
%	\vspace{-0.4cm}
	\caption{CDF (and boxplot with red dot representing the Mean value) of the $N_f$ of users in M1 and M14}
    \label{fig:CDF_popularity_trend_all}
%\vspace{-0.2cm}
\end{minipage}
%\vspace{-0.2cm}
\hspace{0.06cm}
\begin{minipage}[b]{0.463\linewidth}
	\centering
	\includegraphics[width=\textwidth]{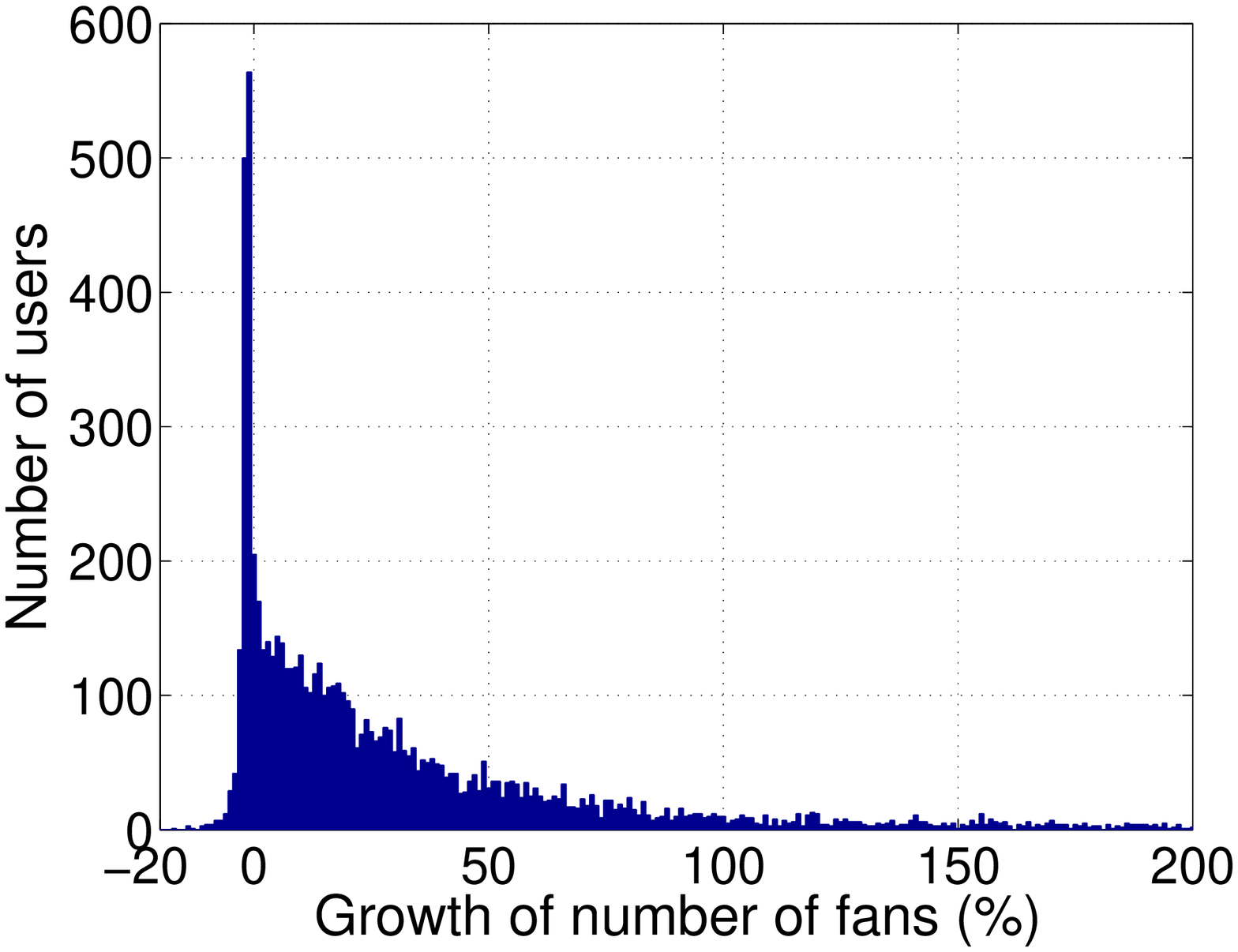}
	\vspace{-0.3cm}
	\caption{Distribution of users based on the percentage of their $N_f$ growth, during 14 months (from M1 to M14)}
%	\vspace{0.1cm}
	\label{fig:hist_growth_percent}
%	\vspace{+0.1cm}
\end{minipage}
\end{figure}

Monthly popularity value is defined indicating the average value of user's $N_f$ in each month.  Since our dataset covers 14 months, each user has a 14-entries vector representing her popularity trend in the period of the dataset.

By considering the overall changes in $N_f$ from M1 to M14 for each user, despite the probable peaks and drops, 80\% (5798 out of 7216) of the users attracted new fans and on the other hand 20\% (1418 out of 7216) lost fans during this 14-month period.
Figure \ref{fig:CDF_popularity_trend_all} shows the distribution of users' popularity from the first month (M1) to the last (M14). The median values for M1 and M14 distributions are 1.3 and 1.7  Millions respectively, which this median value increased from M1 to M14 by 30\% (and 38\% increment for mean value).

Figure \ref{fig:hist_growth_percent} represents the distribution of users based on the percentage of their $N_f$ growth during the period of this study. As shown in the figure, the growth rate of the number of fans for pages who lost fans is not less than -20\% and the major range of fans lost are between -5\% and 0\%.
On the other hand, most of the fan-attractor pages are in the range of 10\% to 30\% growth and the distribution continues in a long-tailed pattern.

\begin{table}[t]
	\centering
	\scriptsize
	\caption{Populated categories distribution in the dataset. Fifth and sixth columns indicate the growth rate of average and the median of $N_f$ over 14 months respectively.}	
\vspace{-0.2cm}
\scalebox{0.95}{
		\begin{tabular}{|c|c|cccc|}
			\hline
			\textbf{\#} & \textbf{FB Category} & \textbf{\#Pages} & \textbf{\%Pages} &\multicolumn{1}{p{1.2cm}}{\textbf{\%Avg. growth}} &\multicolumn{1}{p{1.2cm}|}{\textbf{\%Median growth}} \\
			\hline
			1    & Musician Band & 1231 & \textbf{17} & 47 & 32 \\
			2    & Community & 986 & 13.7  & \textbf{2.1} & \textbf{-1.5} \\
			3    & Tv Show & 477  & 6.6  & 53 & 15 \\
			4    & Movie & 413  & 5.7  & 28 & 18 \\
			5    & Food Beverages & 302  & 4.2  & 19 & 11 \\	
			6    & Product Service & 267  & 3.7 & 24 & 15 \\
			7    & Public figure & 246  & 3.4  & 64 & 33 \\
			8    & Company & 188  & 2.6  & 23  & 15\\
			9    & Athlete & 188  & 2.6  & \textbf{101} & 65\\
			10   & Actor Director & 179  & 2.5  & \textbf{97} & 50\\
			11   & Entertainment & 166 & 2.3 & 26 & 4\\
			12   & App page & 143  & 2.0  & 17 & 8 \\
			13   & Clothing & 139 & 1.9 & 29 & 19\\
			14   & Media News & 134 & 1.8 & 76 & 42\\
			15   & Sports Team & 125  & 1.7  & \textbf{92}  & 60\\
			16   & Games Toys & 109  & 1.5  & 13 & 6 \\
			17   & Health Beauty & 85  & 1.2  & 17 & 7 \\
			\hline
		\end{tabular}%
	}
	\label{tab:fb_categories_top8k}%
\end{table}%

\subsection{Popularity Analysis - Category Wise}
Each page is assigned to a business sector by the page owner in the time of subscribing called category. To investigate the users' distribution and overall popularity evolution inside the categories, 
 we chose 17 (out of 158) categories those that include more than 1\% of the total pages in the dataset separately and more than 75\% in sum shown in Table \ref{tab:fb_categories_top8k}
The main observations from Table \ref{tab:fb_categories_top8k} are as follow:
(i) \textsl{Musician Band} is the most populated category in our dataset which shows users in this category are the most popular ones in the dataset. 
(ii) The percentage of average growth in the fifth column refers to the average $N_f$ growth of users in each category over 14 months. Interestingly, it shows that the \textsl{Athlete, Actor Director}, and \textsl{Sports Team} categories have the highest percentage of growth, and on the contrary \textsl{Community} has the lowest.
This indicates that users in the three mentioned categories are successful in attracting new fans on average, whereas \textsl{Community} category users show a negative growth.
(iii) The last column of the table shows the users' median value of the $N_f$ growth in each category. A negative value here shows users of that category are loosing fans which means people unfollow the pages by \textsl{unliking}. \textsl{Community} is the only category which has negative median growth. This means that most of the users in this category have lost some of their fans.

%%%%%%%%%%%%%%%%%%%%%%%%%%%%%%%%%%%%%%%%%%%%%
\section{Users' Clustering}
\label{sec:Clustering}

This section aims to analyse the popularity in the user level and try to identify different clusters of users with similar patterns in their popularity trends.
To this end, the evolution of $N_f$ is modeled by exploiting different clustering techniques and investigating different characteristics (popularity range, category and activity distributions) in each identified cluster.

\subsection{Feature Vector and Clusters}

  \begin{figure}
	\centering
 	\begin{subfigure}[t]{0.225\textwidth}
  		%\captionsetup{font=scriptsize}
  		\includegraphics[width=\textwidth]{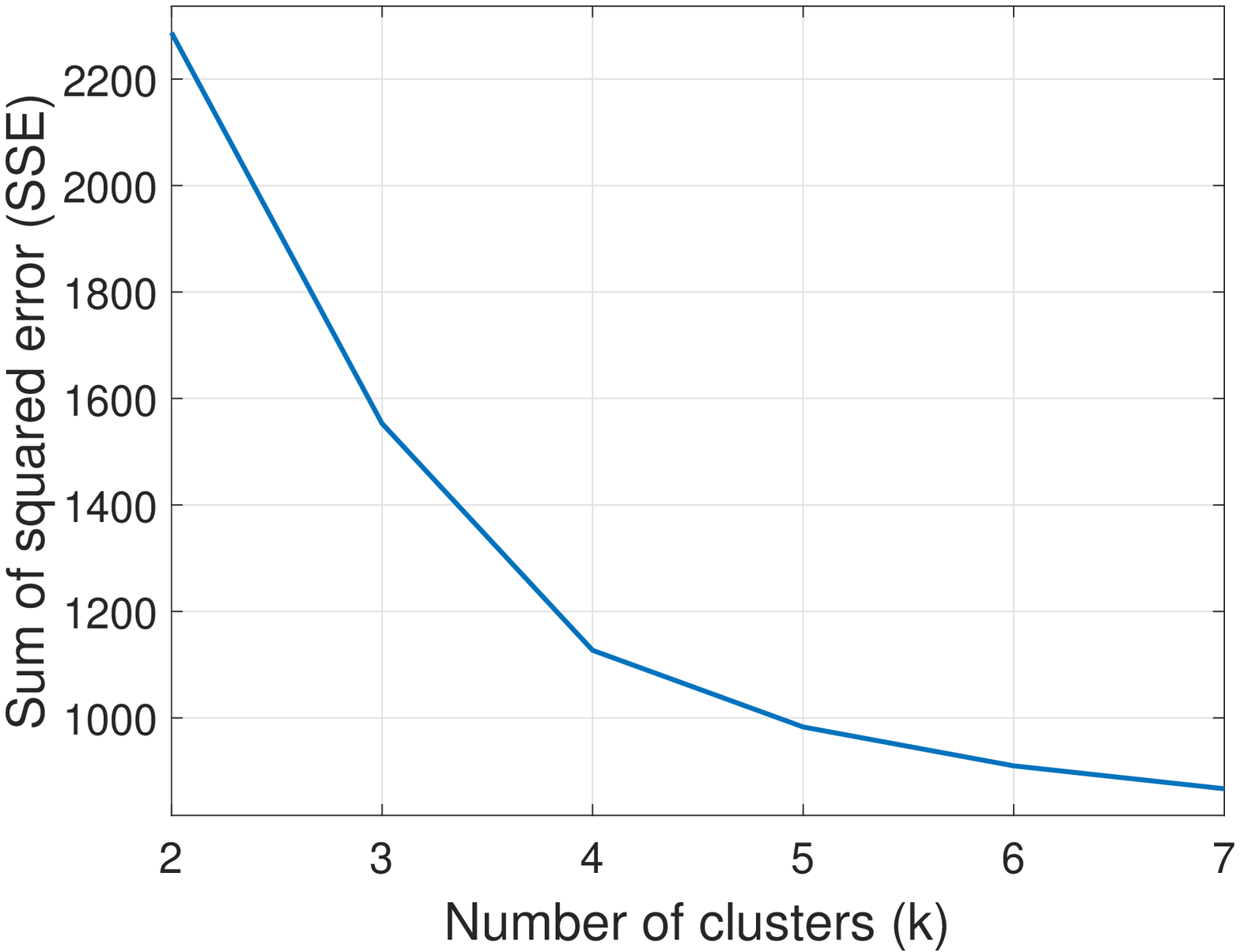}
  		\caption{Sum of squared error (SSE)}
  		\label{fig:sse}
  \end{subfigure}
	~
  \begin{subfigure}[t]{0.225\textwidth}
  		%\captionsetup{font=scriptsize}
  		\includegraphics[width=\textwidth]{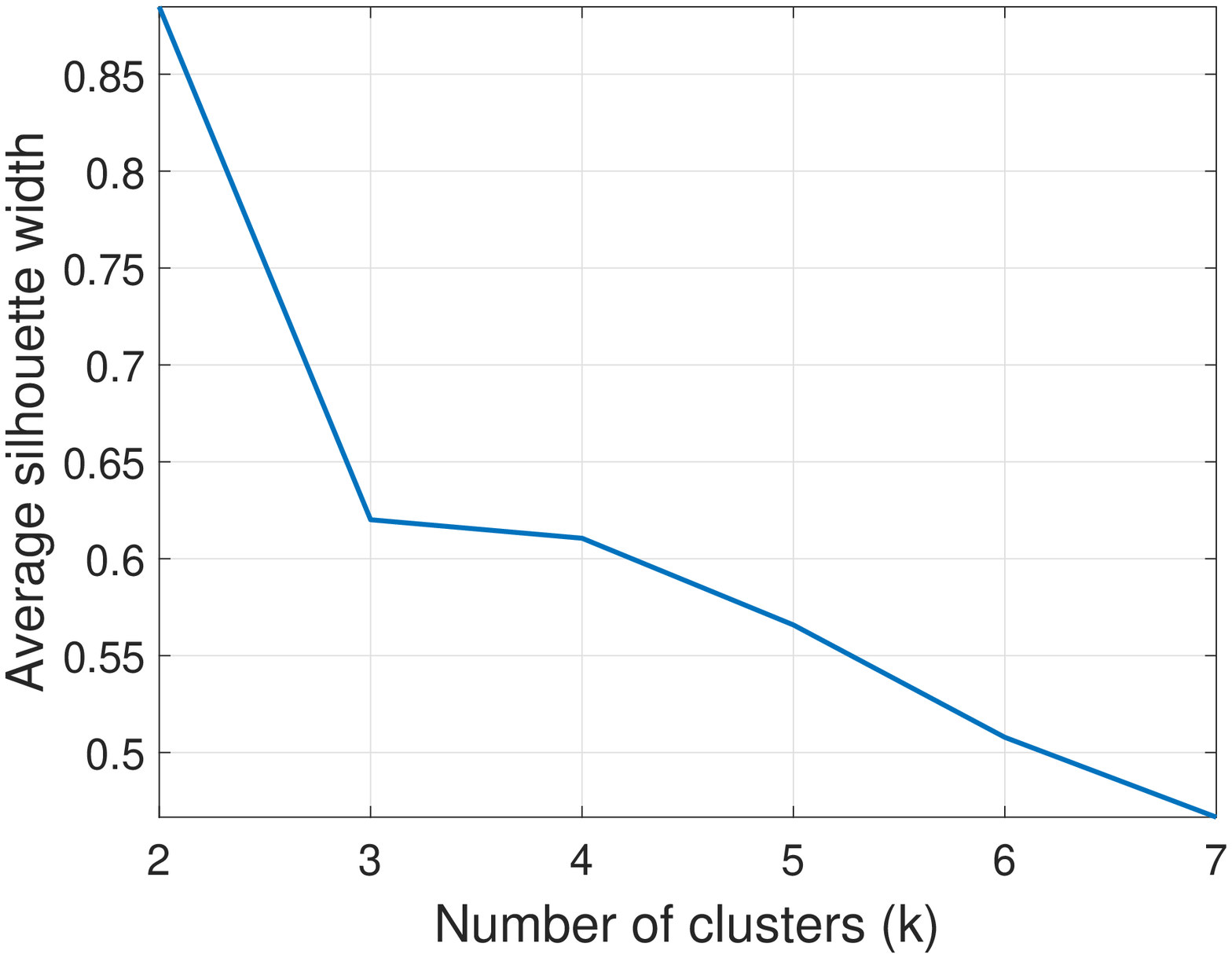}
  		\caption{Silhouette width for different cluster numbers}
  		\label{fig:silhouette}
  	\end{subfigure}
	~
		\vspace{-0.15cm}
  	\caption{SSE test to find the appropriate k value for our dataset}
		\label{fig:choosing_best_k}
  \end{figure}

To cluster users based on the popularity attributes, a 14-entry monthly popularity vector for each user is used as a feature vector in the clustering method. The entries represent the monthly $N_f$ of users that have values over the range of one hundred thousand to one hundred million. The goal is to group the users with similar popularity evolution into a cluster, regardless of the value of $N_f$. To clarify this point, consider two \textsl{FanPages} from quite different ranges of popularity, which both have 50\% growth of $N_f$ with the same trend over the same time period. They should be assigned to a same cluster because their popularity trend are similar. To this end,
we used the Min-Max normalization method which scales every feature vector into $[0,1]$ by obtaining the values 0 and 1 at the minimum and maximum points, respectively.
The feature vectors thus represent the time-series popularity trends of users.

Next we applied several clustering algorithms including K-means \cite{macqueen1967some}, KSC \cite{yang2011patterns} and K-shape \cite{paparrizos2015k} and as the outcome of all of them were similar, we consider the K-means clustering algorithm to the above-mentioned feature vectors.  K-means requires the number of clusters (k) as the input parameter. 
There are different approaches to detect the optimal number of clusters. In this study, we used 
the elbow method \cite{kodinariya2013review}, which considers the within-cluster sum of the squared errors (SSE) to find the appropriate k for our dataset. Figure \ref{fig:sse} shows the SSE results for different k numbers applied to the dataset

As depicted in Figure \ref{fig:sse}, the distortion of SSE goes down rapidly by increment of k to the value of 4. Then it descends slowly to 5 and continues with slower decrement. It seems that the diagram reaches an elbow at $k=4$. However to be more assured of an appropriate k value, the Silhouette width \cite{kaufman2009finding} of different k values is also computed. The concept of silhouette width involves the difference between the within-cluster tightness and the separation from the rest of clusters. 

Figure \ref{fig:silhouette} shows the average Silhouette width for different numbers of cluster.
The average Silhouette width is almost constant with k increasing from 3 to 4. This means that with k equals to 4, users are located in as right cluster as with 3. But as the SSE in Figure \ref{fig:sse} has an impressive decrease with 3 clusters, we chose 4 as the appropriate number of clusters.

\begin{figure}[t]
	\centering
	\includegraphics[width=0.38\textwidth, height=0.28\textwidth]{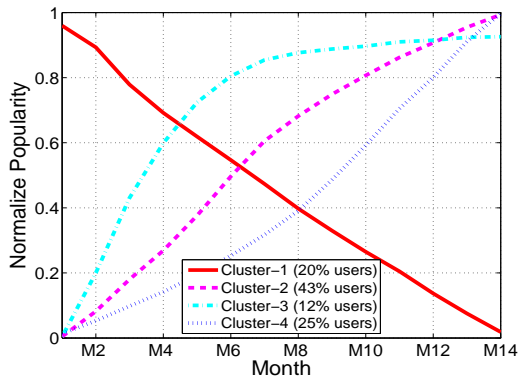}
	\vspace{-0.15cm}
	\caption{Normalized popularity trends of four clusters}
	\label{fig:EM_Trend}
	\vspace{-0.3cm}
\end{figure}

Figure \ref{fig:EM_Trend} represents the normalized popularity trends for the clusters. Each plot shows the average value of the normalized $N_f$ belonging to the users in one of the cluster. In general, three of the identified popularity patterns are ascending by means of different behaviors, and one of them is descending.
In summary we can observe the following points:

\textsl{(i)}
Users are continuously losing their fans in the first cluster (\textsl{Cluster-1}) which includes 20\% of our dataset population.

\textsl{(ii)}
The most populated cluster is the \textsl{Cluster-2} by 43\% of the users. It shows an ascending popularity growth behavior in average. This means that the popularity of the users in this cluster is constantly increasing due to attracting new fans.

\textsl{(iii)}
\textsl{Cluster-3} has 13\% of the dataset population and users in this cluster show a sudden growth (around 80\%) in the first half of the time and then their growth is stopped and somehow saturated in the second half.

\textsl{(iv)}
\textsl{Cluster-4}, with 25\% of the users, shows an opposite behavior to \textsl{Cluster-3}. Its users show near to 30\% growth in the first 7 months and then 70\% during the last 7 months.

Next we characterize the identified clusters from three perspectives, their popularity, category and activity.

\subsection{Popularity Distribution in each Cluster}
\begin{figure}[t]
	\centering
	\includegraphics[width=0.4\textwidth, height=0.33\textwidth]{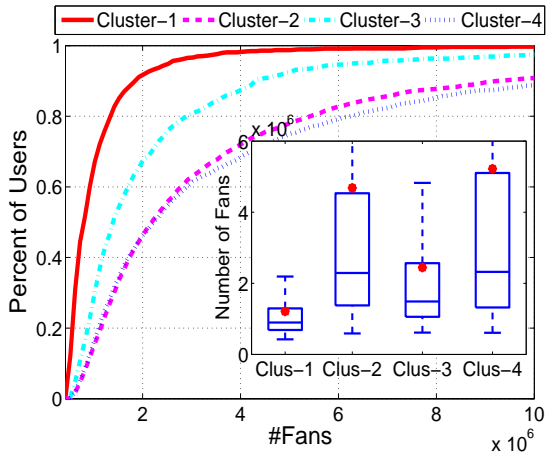}
	\vspace{-0.15cm}
	\caption{CDF (and boxplot) of the distribution of users $N_f$ in four identified clusters. (the red dot inside boxplot represents the Mean value of the distribution).}
	\label{fig:clusters_pop_CDF}
\end{figure}

This section analyzes the clustering results with respect to the users' popularity distribution.
The aim is to identify how the normalized popularity trend can be affected by the absolute value of $N_f$.
First we look to the distribution of popularity in the clusters. Figure \ref{fig:clusters_pop_CDF} shows the CDF plots of the last month (M14) users' popularity in four identified clusters.
The first interesting point in this figure is the popularity distribution of users in \textsl{Cluster-1}. As we saw earlier in Figure \ref{fig:EM_Trend}, users in this cluster are gradually losing their fans. Figure \ref{fig:clusters_pop_CDF} shows most of these users are less popular than the users in other clusters. Almost 65\% of them have less than 1M fans, and the number of users which have more than 2M fans does not exceed 10\%.

According to this plot, three other clusters include users with much higher values of $N_f$. It can be observed that users in two of the most fan-attractor clusters (\textsl{Cluster-2} and \textsl{Cluster-4}) are more popular and have high $N_f$ in compare to users in the other two clusters.
The median values of popularity in these two clusters are almost 2M fans. While only 30\% and 10\% of users in  \textsl{Clusters 3} and  \textsl{1} have more than 2M fans.

Thus, the most popular users belong to \textsl{Cluster-2} and \textsl{Cluster-4}, which both represent exclusively fan-attractor behaviour.
In contrary, most of the less popular users are in \textsl{Cluster-1} and \textsl{Cluster-3}, where their popularity pattern show a fan losing behavior or of being almost saturated.
To conclude this section, in general more popular users show very sharp fan attracting trends while less popular ones show fan losing or saturating trends.

\subsection{Category Analysis}
\begin{figure}[t]
	\centering
	\includegraphics[width=0.49\textwidth, height=0.29\textwidth]{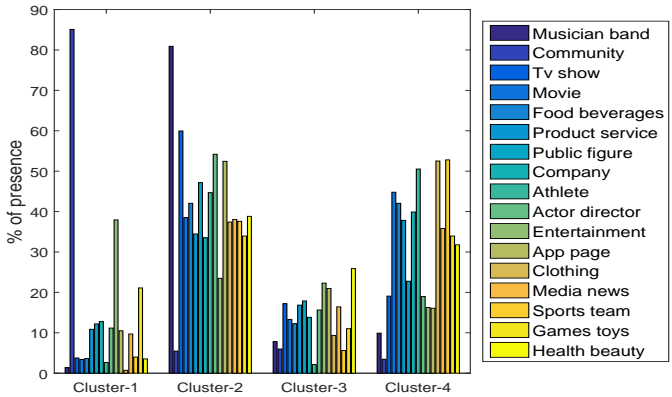}
	\vspace{-0.15cm}
	\caption{Distribution of predefined Facebook categories in each identified cluster}
	\label{fig:clusters_category_distribution}
	\vspace{-0.3cm}
\end{figure}

In this part we investigate the distribution of categories inside the identified clusters to understand if there are categories with a dominant population in a specific cluster.
Figure \ref{fig:clusters_category_distribution} shows the distribution of the 17 most populated categories, mentioned earlier in Table \ref{tab:fb_categories_top8k}, across the identified clusters.

An interesting observation from the category distribution is the high presence of the \textsl{Community} and \textsl{Entertainment} categories in \textsl{Cluster-1}, with around 85\% and 40\% portion of presence, respectively. Given that the users in this cluster are losing their fans, and the \textsl{Community} category is the second most populated category with 13.7\% of the users in the dataset, it can be concluded that it is also the biggest set of fan-loser users. According to the Facebook\footnote{https://www.facebook.com/help/187301611320854/},
``a Community Page is a page about an organization, celebrity or topic that it does not officially represent. It links to the official page about that topic." Our observations show that a \textsl{Community} page is a place that Facebook users gather to share their ideas, images, posts around a specific topic, company, or celebrity and cannot remain attractive to users over time. 
 One of the reason we found is the new feature of Facebook ``Verified" which provide the possibility for verifying popular pages which Facebook started in May 2013. After verification, people are more likely following the verified pages instead of the community pages.

In summary, according to the popularity trend of other three clusters and category distributions of \textsl{Cluster-1}, we can say more than 80\% of users from all categories except \textsl{Community} and \textsl{Entertainment} categories are attracting new fans.

\textsl{Cluster-2}, which shows a fixed rate of popularity growth, includes a high presence of \textsl{Musician band} and \textsl{TV show} categories, which are two of the three most-populated categories with 17\% and 6.6\% of the users in the dataset. These two categories, accompanied by \textsl{Actor director}, contain most of the celebrities' pages 
in our dataset.
 On the other hand, as \textsl{Cluster-2} shows the most successful fan-attracting trend, we can indicate that the pages of celebrities are always interesting for people to follow. Around 30\% to 50\% of other categories' users also show similar pattern of attracting new fans.

The distribution of categories in \textsl{Cluster-3} shows almost an equal presence of all categories without any dominant one, except a minimum presence of \textsl{Athlete} categories. 
The trend of this cluster could have different explanations like fan-saturation, reduction of the activity or external events which have the same side effect on users in different categories. 
In the next section, we look for the effect of activity volume on users' fan-trends as a probable influential factor.

\textsl{Cluster-4}, which includes 25\% of our users, has a variety 
of categories distribution. Three categories, \textsl{Athlete},  \textsl{Clothing}, and \textsl{Sport team} have more than 50\% of their population in this cluster. According to the popularity pattern of this cluster, most of the users experienced more than 70\% of their popularity growth in the second half of the study period. Some famous celebrities such as \textsl{Neymar} (Football player), \textsl{Real Madrid C.F.} (Sport team) are in this cluster. For users such as those related to football, the most probable reason of significant $N_f$ growth may be the main events of European leagues which are overlapped with the second half of our dataset period.

As a summary of this part, we saw that \textsl{Community} is characterized as the most fan-losing category with a major presence in \textsl{Cluster-1}. The categories containing more celebrities are the most fan-attracting ones, with a significant presence in the two most fan-attractor clusters, \textsl{Cluster-2} and \textsl{Cluster-4}. 

\subsection{Activity Analysis}
\label{activity_cluster}
\begin{figure}
	\centering
	\begin{subfigure}[t]{0.23\textwidth}
		%\captionsetup{font=scriptsize}
		\includegraphics[width=\textwidth]{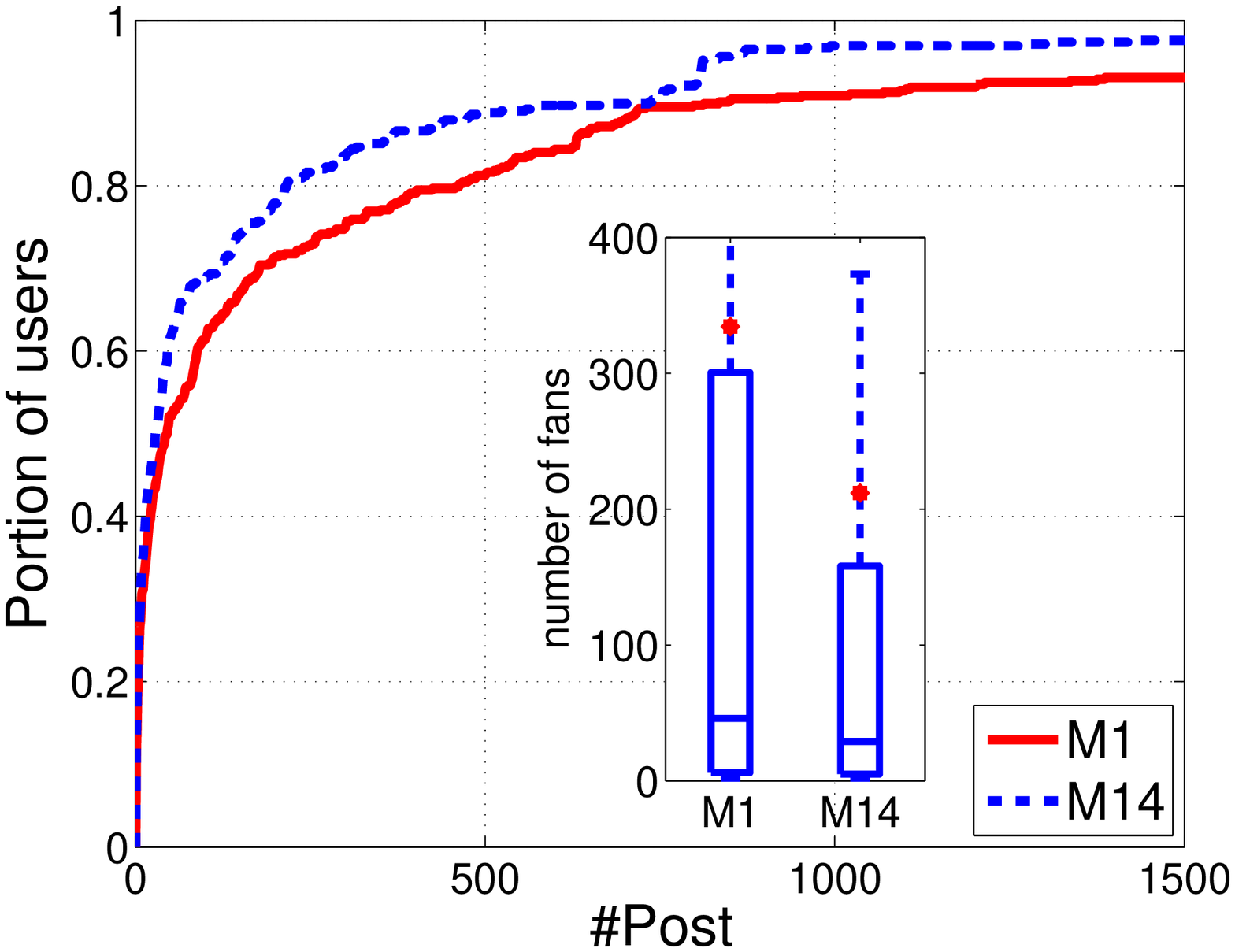}
		\vspace{-0.5cm}
		\caption{Cluster-1}
		\label{fig:cluster1}
		%\vspace{-0.3cm}
	\end{subfigure}
	~
	\begin{subfigure}[t]{0.23\textwidth}
		%\captionsetup{font=scriptsize}
		\includegraphics[width=\textwidth]{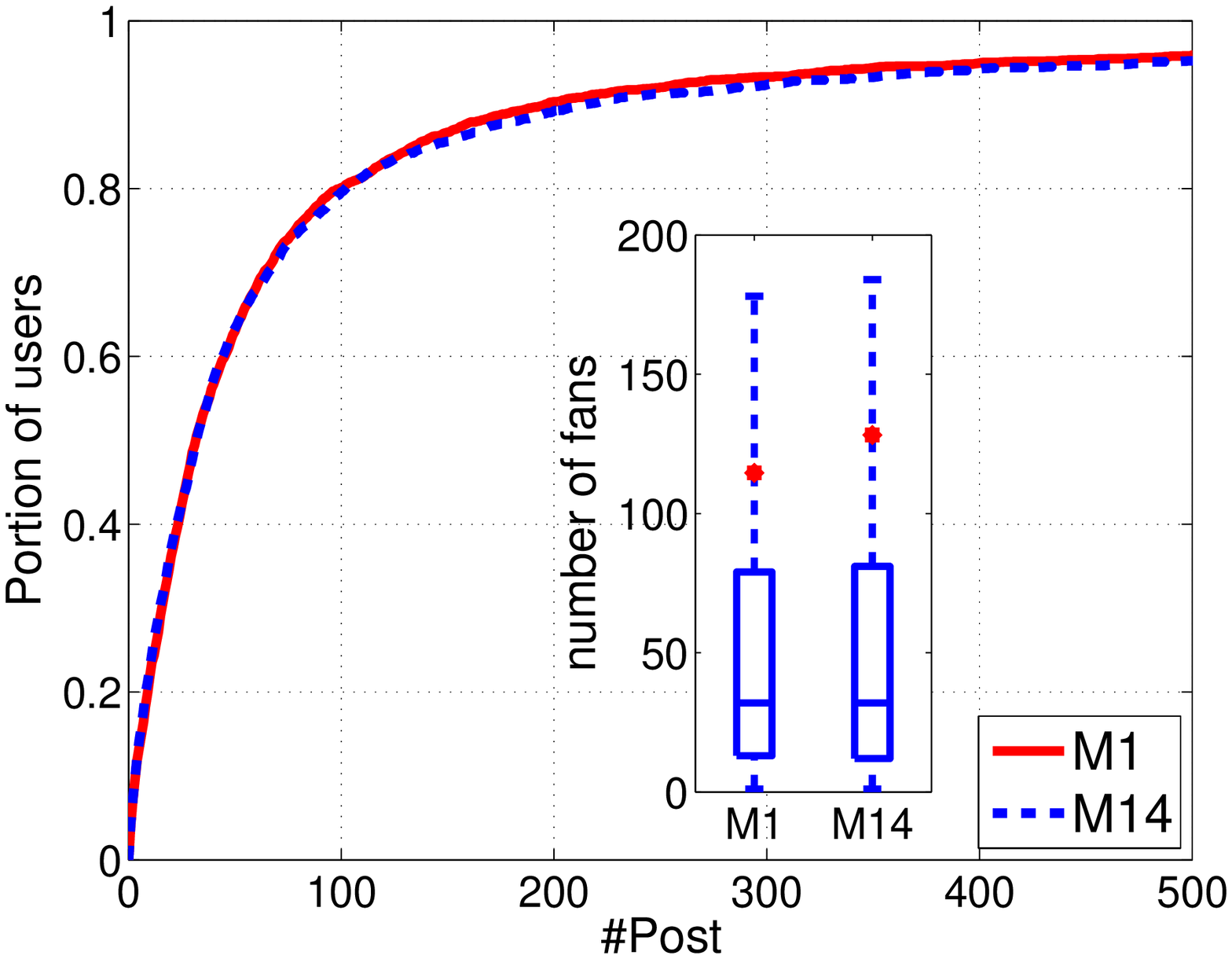}
		\vspace{-0.5cm}
		\caption{Cluster-2}
		\label{fig:cluster2}
		%\vspace{-0.3cm}
	\end{subfigure}
	%~
	\\
	%\hspace{0.1cm}
	\begin{subfigure}[t]{0.227\textwidth}
		%\captionsetup{font=scriptsize}
		\includegraphics[width=\textwidth]{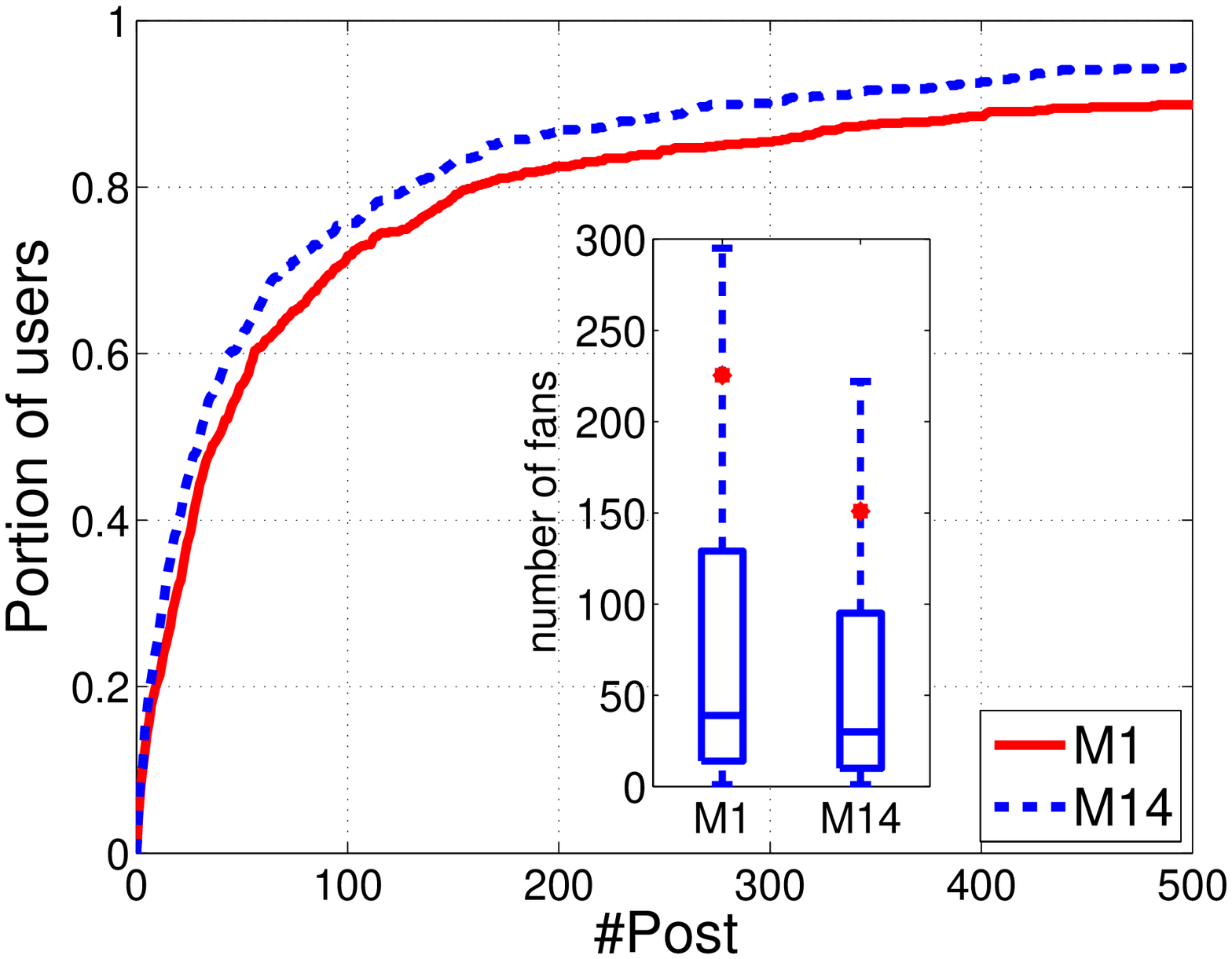}
		\vspace{-0.5cm}
		\caption{Cluster-3}
		\label{fig:cluster3}
		%\vspace{-0.3cm}
	\end{subfigure}
	~
	\begin{subfigure}[t]{0.227\textwidth}
		%\captionsetup{font=scriptsize}
		\includegraphics[width=\textwidth]{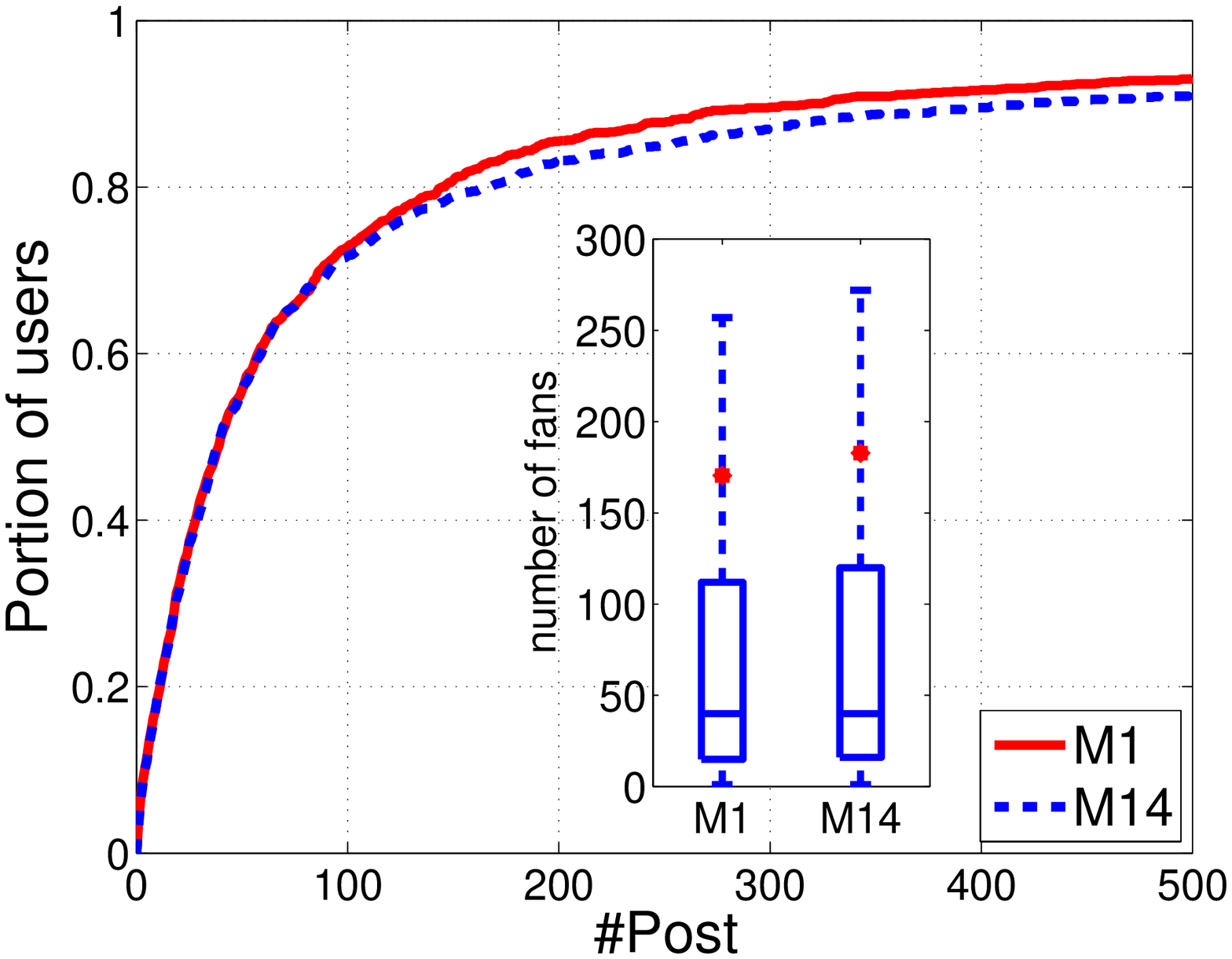}
		\vspace{-0.45cm}
		\caption{Cluster-4}
		\label{fig:cluster4}
		%\vspace{-0.3cm}
	\end{subfigure}
		~
	\vspace{-0.2cm}
	\caption{CDF (and BoxPlot) of number of published posts per user in the first (M1) and last (M14) months of the dataset (red dot in boxplot represents the Mean value of the distribution).}
	\label{fig:CDF_activity_first_last}
	\vspace{-0.2cm}
\end{figure}

Being active in Facebook by continuously publishing new posts, can ensure professional users to stay in touch with their followers and attract new ones as well \cite{leung2013attracting}. 
To understand the impact of activity on popularity, Figure \ref{fig:CDF_activity_first_last} shows the CDF plots of the number of published posts by users in four clusters for M1 and M14. It illustrates that the published posts of the users in \textsl{Cluster-1}, who lost their fans, declined from M1 to M14. 
This can be observed for the distribution of users in \textsl{Cluster-3} as well (Figure \ref{fig:cluster3}). As discussed before, the $N_f$ of users in this cluster is almost constant for the second half of the study period. It can be concluded that the reduced number of activity in these two clusters is an important factor for the lost of fans in \textsl{Cluster-1} and the failure to attract new ones in \textsl{Cluster-3}.

In contrast, the activity level of users have not changed substantially in the two most fan-attracting clusters, \textsl{Cluster-2} and \textsl{Cluster-4}. Even we can see a small increment in the activity curve of \textsl{Cluster-4}; the number of users who published more than 150 posts in the last month is greater than the number of users who posted that much in the first month. Considering their popularity trends which show a continuous growth, it can be deducted that being constantly active effect the process of attracting new fans.

In a nutshell, we observe that staying active in terms of publishing posts can help to attract new fans and followers whereas reducing the activity level can lead to stagnant number of followers, and even losing fans.

\section{Conclusion}
\label{sec:Conclusion}
This paper studied the users popularity evolution in online social networks with a focus on professional users such as companies, celebrities, brands, and etc. To this end, the number of fans of almost 8K of the most popular professional users was collected in six daily snapshots, over a period of 14 months.
The users' published posts were also collected in the same time period, which eventually provided around 20 million snapshots of popularity values. 
The experiments conducted on this data reveal interesting results.
Users were categorized into two main groups fan-losers and fan-attractors, and four different patterns of popularity evolution were identified.
Several factors are identified that influence the popularity trend of users, such as the social position like celebrities, external events associated to the owner of the page, and the level of activity. 
 The findings from this study provide a comprehensive view on professional users' popularity evolution, and reveal the impact of different factors on it.

This study only analyzed professional Facebook users. The analysis of cross-popularity of these users on other major social networks, e.g. Twitter, Instagram, etc., can be considered as a future work.
Beside the activity and external events, it could be very interesting to look on other potential influential factors such as specific strategies that users are following in social media.
Providing a comprehensive list of suggestions for users to enhance their success in social media can also be an extension of this work.

\bibliographystyle{IEEEtran}
\bibliography{citation}

\end{document}